\documentclass[12pt]{article}

\usepackage{newtxtext,newtxmath}
\usepackage[left]{lineno}

\usepackage{graphicx}

\usepackage[letterpaper,margin=1in]{geometry}

\renewenvironment{abstract}
	{\quotation}
	{\endquotation}

\date{}

\makeatletter
\renewcommand{\fnum@figure}{\textbf{Figure \thefigure}}
\renewcommand{\fnum@table}{\textbf{Table \thetable}}
\makeatother

\usepackage{scicite}

\usepackage{url}

\def\scititle{
	Advancing global sea ice prediction capabilities using a fully-coupled climate model with integrated machine learning
}

\title{\bfseries \boldmath \scititle}

\author{
	William Gregory$^{1\ast}$,
	Mitchell Bushuk$^{2}$,
	Yong-Fei Zhang$^{3}$,
    Alistair Adcroft$^{1}$,\and
    Laure Zanna$^{4,5}$, 
    Colleen McHugh$^{6}$,    
    Liwei Jia$^{2}$ \and
	\small$^{1}$Atmospheric and Oceanic Sciences Program, Princeton University, Princeton, NJ, 08540, USA.\and
	\small$^{2}$NOAA, Geophysical Fluid Dynamics Laboratory, Princeton, NJ, 08540, USA.\and
    \small$^{3}$Earth System Science Interdisciplinary Center, University of Maryland, Maryland, 20740, USA.\and
    \small$^{4}$Courant Institute of Mathematical Sciences, New York University, New York, 10012, USA.\and
    \small$^{5}$Center for Data Science, New York University, New York, 10011, USA. \and
    \small$^{6}$Science Applications International Corporation, Reston, VA, 20190, USA. \and
	\small$^\ast$Corresponding author. Email: wg4031@princeton.edu
}

\begin{document} 

\maketitle

\begin{abstract} \bfseries \boldmath 
We showcase a hybrid modeling framework which embeds machine learning (ML) inference into the GFDL SPEAR climate model, for online sea ice bias correction during a set of global fully-coupled 1-year retrospective forecasts. We compare two hybrid versions of SPEAR to understand the importance of exposing ML models to coupled ice-atmosphere-ocean feedbacks before implementation into fully-coupled simulations: Hybrid\textsubscript{CPL} (with feedbacks) and Hybrid\textsubscript{IO} (without feedbacks). Relative to SPEAR, Hybrid\textsubscript{CPL} systematically reduces seasonal forecast errors in the Arctic and significantly reduces Antarctic errors for target months May--December, with $>$2x error reduction in 4--6-month lead forecasts of Antarctic winter sea ice extent. Meanwhile, Hybrid\textsubscript{IO} suffers from out-of-sample behavior which can trigger a chain of Southern Ocean feedbacks, leading to ice-free Antarctic summers. Our results demonstrate that ML can significantly improve numerical sea ice prediction capabilities and that exposing ML models to coupled ice-atmosphere-ocean processes is essential for generalization in fully-coupled simulations.
\end{abstract}


\section*{Introduction}
\noindent Over the past 4--5 decades, remote sensing observations, ground-based instruments, and submarine surveys have shown that the Earth's sea ice cover is undergoing dramatic changes. The Arctic, for example, has seen basin-wide thinning and retreat of sea ice across all seasons \cite{Stroeve2018,Kwok2018}. This ice loss has played a significant role in high-latitude climate feedbacks and Arctic amplification, where Arctic surface temperatures have warmed at nearly four times the rate of the global average \cite{Rantanen2022}. Furthermore, Arctic sea ice loss can also contribute to a slow-down in the poleward transport of warm ocean waters \cite{Sevellec2017} and increased frequency of extreme weather events across Europe \cite{Bailey2021} and North America \cite{Barnes2015,Cohen2021}. Meanwhile, Antarctic sea ice area exhibited a modest positive trend between 1979--2014. However, since 2016 there have been five record low summer minima and two record low winter maxima, with many studies now suggesting a regime shift in Antarctic sea ice caused by Southern Ocean warming   \cite{Eayrs2021,Fogt2022,Turner2022,Raphael2022,Purich2023}.

Reproducing these historical sea ice changes within climate models is critical for enabling confident assessments of how future anthropogenic sea ice changes will impact climate and society. Meanwhile, the latest generation of climate models submitted to the sixth phase of the Coupled Model Intercomparison Project (CMIP6) show considerable spread in their simulations of historical sea ice area and trends, with models generally under-estimating the sensitivity of sea ice to global warming in the Arctic \cite{Notz2020} and over-estimating the sensitivity in the Antarctic \cite{Roach2020}. While internal climate variability certainly plays a role \cite{Ding2019,Singh2019}, this spread in CMIP6 sea ice mean state and sensitivities arises primarily due to uncertainty in component and coupled model physics, which in turn drive model-dependent biases \cite{Bonan2021}.

On shorter timescales, these model physics errors also impact our ability to make accurate seasonal-to-interannual sea ice predictions, as models struggle to faithfully reproduce various physical drivers of regional sea ice variability \cite{Guemas2016a,Gregory2022,Bushuk2022}. Since 2008 there has been a growing community effort to understand and improve sea ice prediction capabilities. This effort culminates each year into a ``Sea Ice Outlook'', where community members submit seasonal forecasts of the September Arctic sea ice minimum and February Antarctic sea ice minimum to the Sea Ice Prediction Network (SIPN) online platform \cite{Blanchard2023,Massonnet2023}. Forecasts range from statistical techniques \cite{Schroder2014,Petty2017,Gregory2020} to fully-coupled dynamical models \cite{Johnson2019,Molod2020,Zhang2022}, as well as heuristic approaches. A recent inter-comparison of 34 individual forecast systems that are routinely submitted to SIPN found that many statistical and dynamical models can skillfully predict September Arctic sea ice conditions 1--3 months in advance \cite{Bushuk2024}, suggesting that useful real-time predictions of September Arctic sea ice are likely on the horizon. Meanwhile in a separate SIPN-South inter-comparison study of Antarctic forecasts, statistical models were found to generally outperform coupled climate models at predicting regional-scale sea ice variability \cite{Massonnet2023}. This therefore prompts an urgent need to improve Antarctic sea ice forecasts within climate models.

Achieving useful seasonal-to-interannual climate model sea ice predictions means addressing both the model physics errors which lead to systematic bias and also ensuring accurate initial conditions for the land, atmosphere, ocean and sea ice. Accurate initial conditions are routinely achieved through frameworks such as nudging \cite{Saha2014,Li2021,Lu2020} and data assimilation \cite{Zhang2007,Zhang2021,Zhang2023}. Within which, model states are either linearly relaxed toward a set of observations over a given time window (nudging), or updated through a Bayesian treatment of model and observational uncertainty (data assimilation). In this present study, we investigate specifically the model physics problem, while also utilizing data assimilation to characterize model errors.

The recent growth in application of machine learning (ML) techniques to climate research has been extraordinary. For sea ice, this has led to breakthroughs in remote sensing and sea  ice reanalysis \cite{Dawson2022,Zampieri2023,Au2024,Gregory2024b}, statistical forecasting \cite{Chi2017,Andersson2021}, and has also paved a new-era in ``hybrid'' sea ice modeling---using ML to replace or improve certain aspects of sea ice model physics \cite{Finn2023,Driscoll2024,Gregory2023,Gregory2024a}. Of course, hybrid modeling is not just limited to sea ice, but has been a burgeoning area of research in both atmosphere \cite{Brenowitz2019,Shamekh2023,Mansfield2024,Chapman2025,Watt2024} and ocean \cite{Zanna2020,Sane2023,Subel2023,Perezhogin2024} models as well. One branch of hybrid climate modeling in particular focuses on learning state-dependent representations of structural model error. In this approach, it is assumed that the corrections, or increments, applied to a numerical simulation during data assimilation (DA) or nudging are largely a manifestation of predictable errors associated with poorly parameterized/missing physics and the discretization of continuous equations \cite{Palmer2011}. An ML model can therefore be used to predict these increments using only model state variables as inputs, thus providing a framework for online bias correction during subsequent numerical simulations. This approach has been shown to successfully reduce systematic model biases when run in component and idealized models \cite{Brajard2021,Watt2021,Gregory2024a,Chapman2025}. A recent study also extended this approach to bias correct sea ice and ocean conditions in the fully-coupled Norwegian Climate Prediction Model \cite{He2025}. While their study showed promising bias improvements, their implementation was restricted to the Arctic domain and used different ML models for each prediction month and year, resulting in 236 different ML models. Their approach also only performed bias correction once per roughly 15 days.

In this present article we assess the ability of a novel ML-based sea ice bias-correction framework to improve global sea ice prediction skill in a set of 1-year fully coupled retrospective forecast (reforecast) experiments, using the Geophysical Fluid Dynamics Laboratory (GFDL) seasonal-to-decadal prediction model, SPEAR \cite{Delworth2020}. We not only investigate how this hybrid modeling framework impacts forecast skill, but also pay close attention to the importance of exposing ML models to coupled ice-atmosphere-ocean feedbacks before their implementation into fully-coupled numerical simulations. Building on past work, we bias correct \emph{global} sea ice conditions using a single ML model which takes spatially local inputs of model state variables to predict sea ice concentration (SIC) DA increments at each model grid point \cite{Gregory2023,Gregory2024a}. Furthermore, bias correction is applied at the thermodynamic timestep of the sea ice model and the same ML model is used for all initialization dates and forecast years (see Materials and Methods). To our knowledge, this is the first example of a large-scale sea ice model with fully-integrated ML inference applied to global fully-coupled sea ice forecasts.

\begin{figure}[t!] 
	\centering
	\includegraphics[width=\textwidth]{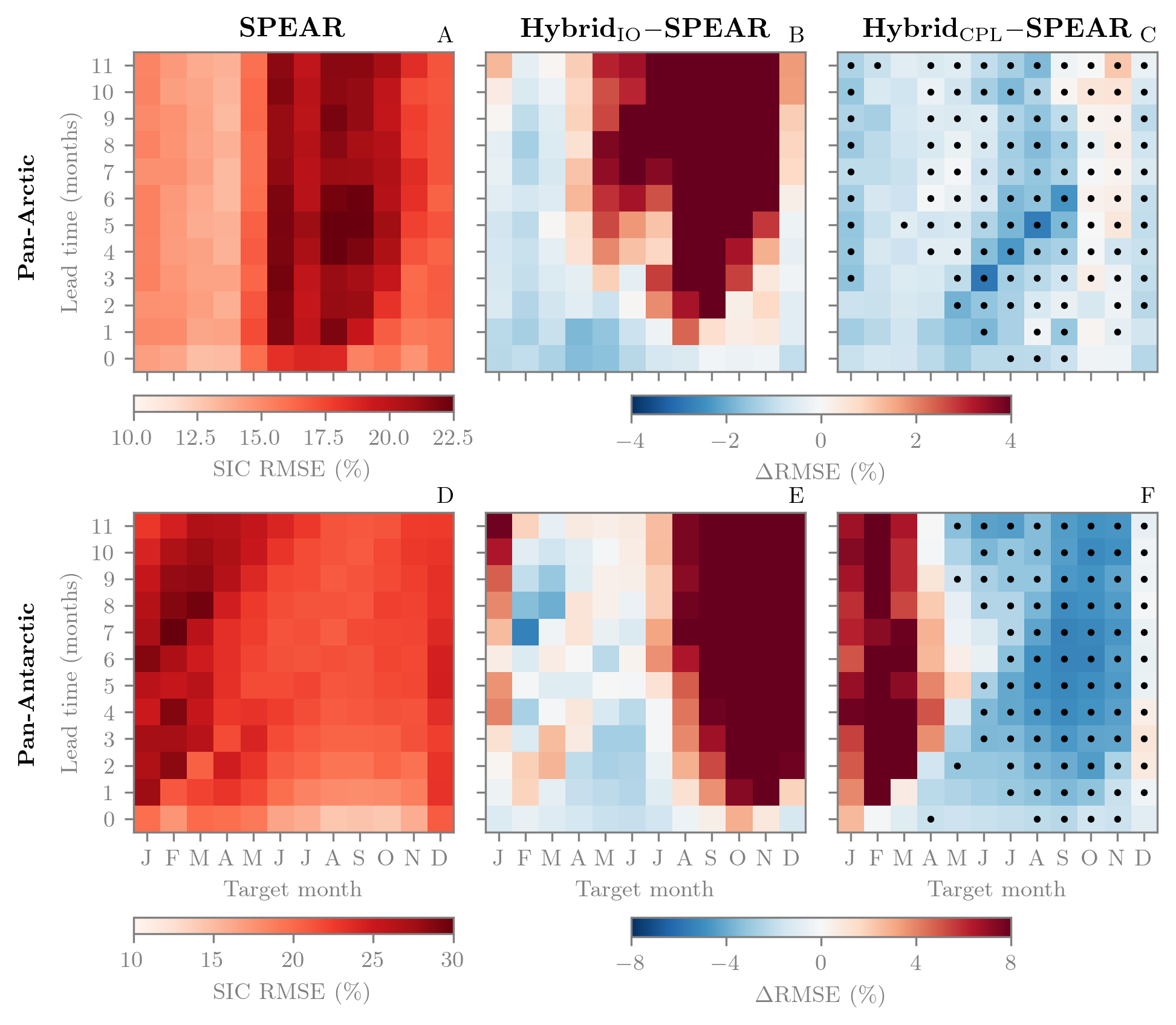}
	\caption{\textbf{Sea ice concentration (SIC) prediction error, 2018--2024.}
		(\textbf{A}) SPEAR pan-Arctic RMSE. (\textbf{B}) RMSE difference between Hybrid\textsubscript{IO} and SPEAR. (\textbf{C}) Same as (\textbf{B}) but for  Hybrid\textsubscript{CPL}. (\textbf{D}--\textbf{F}) Same as for (\textbf{A}--\textbf{C}) but for pan-Antarctic. Stippling in (\textbf{C},\textbf{F}) show where Hybrid\textsubscript{CPL} outperforms Hybrid\textsubscript{IO} at the 95\% confidence level. Errors are relative to NSIDC NASA Team observations \cite{DiGirolamo2022}.}
	\label{fig:forecastskill} 
\end{figure}

\section*{Results}
The 1-year ensemble-based reforecasts in this study are initialized on the first day of each month, for all months between January 2018 and December 2023, providing 72 ensemble forecasts to evaluate in both hemispheres. We compare two different ML models for bias correction in separate suites of reforecast experiments: (i) following \cite{Gregory2024a}, an ML model which is trained to predict SIC DA increments from a 36-year (1982--2017) reanalysis-forced ice-ocean configuration of SPEAR with sea-surface temperature (SST) nudging, (ii) an ML model which is trained to predict SIC DA increments from a fully-coupled configuration of SPEAR which nudges SSTs and the 3D atmosphere temperature, wind, and humidity fields (also between 1982--2017). The sea ice fields in these training data are therefore heavily constrained by both the nudged ocean and the prescribed and nudged atmosphere, respectively. However, the nudged atmosphere configuration allows for coupled ice-atmosphere-ocean feedbacks, whereas the prescribed atmosphere does not. Comparison of hybrid models (i) and (ii), hereafter Hybrid\textsubscript{IO} and Hybrid\textsubscript{CPL}, respectively, will therefore allow us to determine the importance of these coupled feedbacks when training ML models for implementation into free-running (no nudging) fully-coupled simulations. All results in this section are based on ensemble mean fields.

\subsection*{Evaluation of forecast errors}
Fig. \ref{fig:forecastskill} shows the root mean squared error (RMSE) of Arctic and Antarctic SIC predictions, for each target and initialization month. Here, RMSE is computed over all ice-covered grid points in both observations and models for each day of a given target month, and then averaged to produce a monthly-mean value. A lead 0 prediction then corresponds to a forecast initialized on the first day of a given month, to forecast all days in that same month---a January 1 initialized forecast of January, for example. A lead 1 prediction of January is then a forecast initialized on December 1, and so on. For SPEAR (Figs. \ref{fig:forecastskill}A and \ref{fig:forecastskill}D), the RMSE is highest for summer target months in both hemispheres, although the Antarctic generally displays higher year-round RMSE (note the different color bars). Larger summertime errors are expected given that ice melting causes local SIC variations throughout the interior ice pack, while in winter the interior ice pack is predominantly fully ice covered in both observations and models. Figs. \ref{fig:forecastskill}B and \ref{fig:forecastskill}E then show the difference in RMSE between Hybrid\textsubscript{IO} and SPEAR, where blue colors indicate an improved forecast relative to SPEAR and red colors a poorer forecast. The degradation in May--November Arctic predictions and July--January Antarctic predictions are clearly the most striking features of these panels. Meanwhile, Arctic predictions in Hybrid\textsubscript{CPL} are near systematically improved compared to SPEAR (Fig. \ref{fig:forecastskill}C), and are significantly improved (95\% confidence, estimated by a 10,000 sample bootstrapping with replacement) in 72\% of cases over Hybrid\textsubscript{IO}. Between May--December, Hybrid\textsubscript{CPL} systematically reduces Antarctic RMSE relative to SPEAR and Hybrid\textsubscript{IO}. However, forecasts are degraded between January--April. Overall, Hybrid\textsubscript{CPL} shows improvement over Hybrid\textsubscript{IO} in 56\% of cases in the Antarctic, although with only 7 years of validation these values could be subject to internal variability.

\begin{figure}[t!] 
	\centering
	\includegraphics[width=\textwidth]{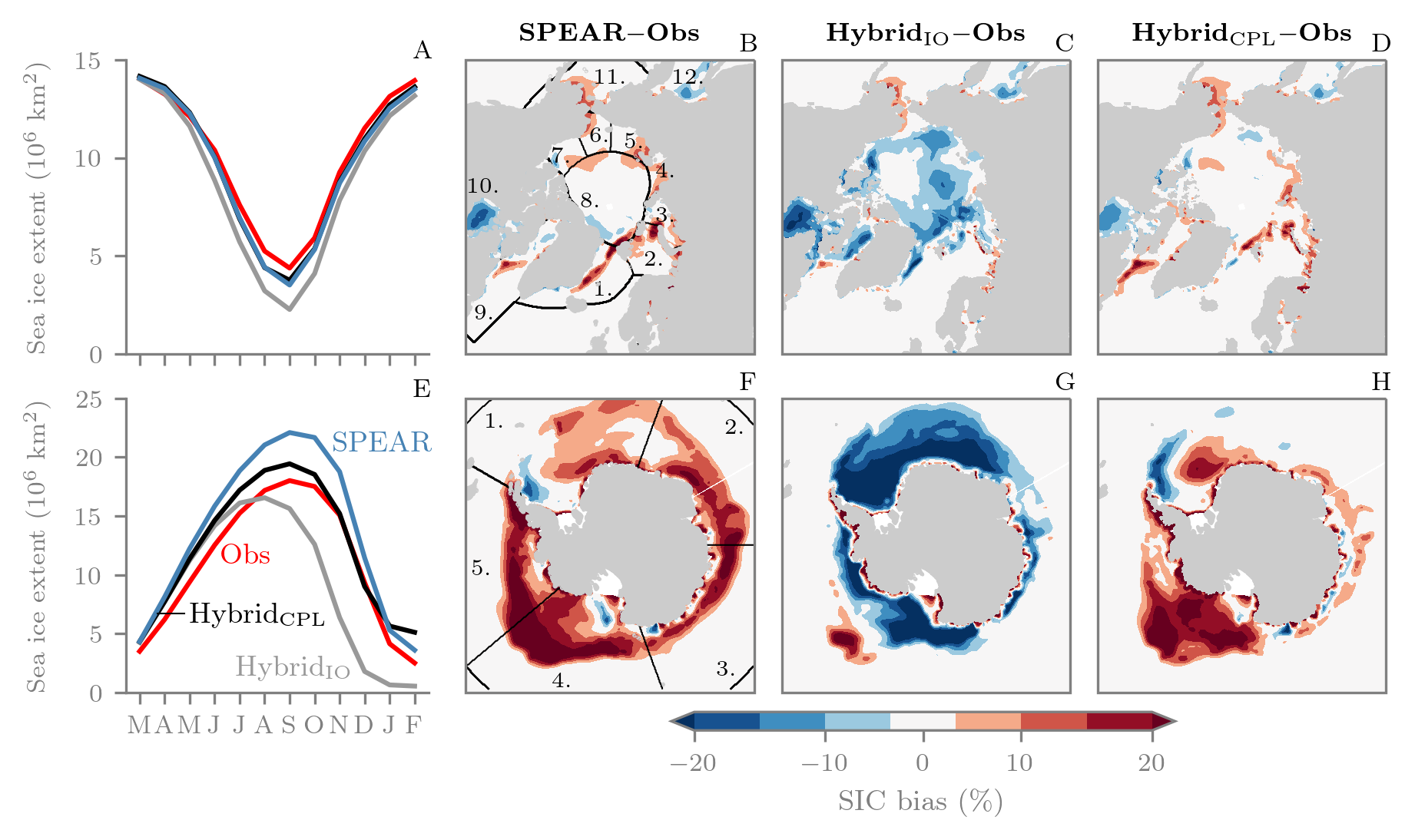}
	\caption{\textbf{March-initialized reforecast bias, 2018--2024.}
		(\textbf{A}) Mean pan-Arctic sea ice extent. (\textbf{B}--\textbf{D}) Sea ice concentration (SIC) bias across entire 1-year reforecasts for SPEAR, Hybrid\textsubscript{IO} and Hybrid\textsubscript{CPL}, respectively. (\textbf{E}--\textbf{H}) Same as (\textbf{A}--\textbf{D}) but for Antarctic. Biases are relative to NSIDC NASA Team observations \cite{DiGirolamo2022}. Regions in (\textbf{B}) are 1. GIN Sea, 2. Barents Sea, 3. Kara Sea, 4. Laptev Sea, 5. East Siberian Sea, 6. Chukchi Sea, 7. Beaufort Sea, 8. Central Arctic, 9. Baffin Bay and Labrador Sea, 10. Hudson Bay, 11. Bering Sea, 12. Sea of Okhotsk. Regions in (\textbf{F}) are 1. Weddell Sea, 2. Indian Ocean, 3. Pacific Ocean, 4. Ross Sea, 5. Amundsen and Bellingshausen Sea.}
	\label{fig:MarchBias} 
\end{figure}

To take a closer look at the performance of each model, Fig. \ref{fig:MarchBias} now shows March-initialized reforecast biases. Starting with the Arctic, the baseline SPEAR model generally performs well at Arctic sea ice forecasts, ranking 2\textsuperscript{nd} against 16 other dynamical model predictions of September Arctic sea ice in a recent inter-comparison \cite{Bushuk2024}. In Fig. \ref{fig:MarchBias}A we can see that SPEAR (blue line) tracks the observed pan-Arctic sea ice extent (red line) well from March--June, although starts to diverge in July and slightly under-predicts the September minimum. The mean SIC error across the 1-year reforecasts (Fig. \ref{fig:MarchBias}B) then shows that SPEAR generally has too much sea ice in places such as the Greenland, Iceland, Norwegian (GIN), Barents, Laptev, Chukchi, and Bering seas, and too little sea ice in Hudson Bay and the Sea of Okhotsk. Figure \ref{fig:MarchSPEAR} shows a breakdown of the SPEAR reforecast biases month-by-month, highlighting that the summertime under-prediction appears to originate in Hudson Bay and Baffin Bay in June, and then spreads to the Canadian Archipelago by August and September. Meanwhile, the Hybrid\textsubscript{IO} sea ice extent (gray line) starts to diverge from observations in May (Fig. \ref{fig:MarchBias}A), resulting in systematic under-prediction for the remainder of the forecast period. Fig. \ref{fig:MarchBias}C then shows that Hybrid\textsubscript{IO} has over-corrected the majority of SPEAR's positive SIC biases, with now predominantly negative SIC biases relative to observations. For Hybrid\textsubscript{CPL}, pan-Arctic extent is largely overlapping with SPEAR (Fig. \ref{fig:MarchBias}A, black line), however  displays improved performance for local SIC predictions. Some noteworthy features in the spatial bias map (Fig. \ref{fig:MarchBias}D) include an almost eradication of a systematic GIN Sea bias, along with improvements in Hudson Bay, the East Siberian Sea, the Beaufort Sea, and the Sea of Okhotsk.

\begin{figure}[t!] 
	\centering
	\includegraphics[width=\textwidth]{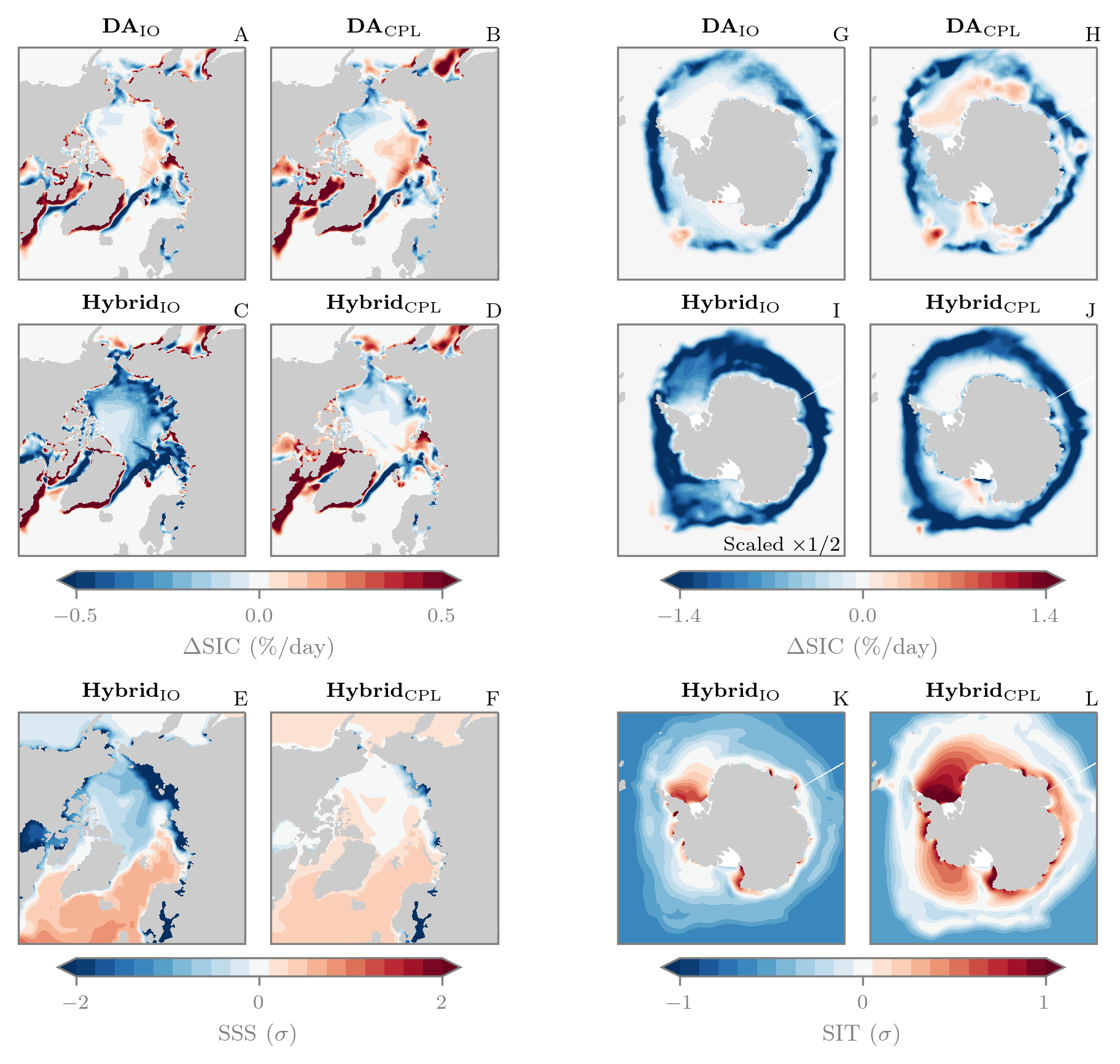}
	\caption{\textbf{Ice-ocean (IO) and coupled (CPL) sea ice concentration (SIC) increments and ML inputs.}
		(\textbf{A}) Mean March--July SIC increments from data assimilation between 1982--2017, from the reanalysis-forced IO simulation. (\textbf{B}) Same as (\textbf{A}) but for the CPL simulation with atmospheric nudging. (\textbf{C},\textbf{D}) Mean 2018--2024 March--July machine learning increments from March-initialized reforecasts with Hybrid\textsubscript{IO} and Hybrid\textsubscript{CPL}, respectively. (\textbf{E},\textbf{F}) Same as (\textbf{C},\textbf{D}) but for normalized sea-surface salinity. (\textbf{G}--\textbf{J}) Same as (\textbf{A}--\textbf{D}) but for Antarctic June--August. (\textbf{K},\textbf{L}) Same as (\textbf{I},\textbf{J}) but for normalized sea ice thickness.}
	\label{fig:increments} 
\end{figure}

Turning to the Antarctic (Fig. \ref{fig:MarchBias}E), SPEAR has a systematic circumpolar year-round positive sea ice extent bias, which is largest in austral winter and has particularly significant contributions from places such as the Ross, Amundsen, and Bellingshausen seas (Fig. \ref{fig:MarchBias}F). For Hybrid\textsubscript{IO}, pan-Antarctic sea ice extent reaches its maximum in August, a full month earlier than both observations and SPEAR (Fig. \ref{fig:MarchBias}E). From this point on, the sea ice extent declines until reaching practically ice-free conditions by February---exemplified by the near hemisphere-wide negative SIC bias in Fig. \ref{fig:MarchBias}G. For Hybrid\textsubscript{CPL}, pan-Antarctic sea ice extent is improved relative to SPEAR in all months except February (Fig. \ref{fig:MarchBias}E). The positive extent bias in winter is dramatically reduced and the early melt season extent (October--December) tracks the observations very well. The spatial bias plot (Fig. \ref{fig:MarchBias}H) highlights notable  improvement in locations such as the Indian and Pacific sectors, the Weddell Sea, as well as the Amundsen and Bellingshausen seas. However, there are some degradations in the southern Ross Sea.

\subsection*{Component vs coupled training}
To understand why Hybrid\textsubscript{IO} systematically under-predicts sea ice conditions in the Arctic and produces near ice-free Antarctic summers, we first investigate potential out-of-sample issues. Starting with the Arctic, we look at the March--July period where the March-initialized Hybrid\textsubscript{IO} reforecasts start to diverge from SPEAR and Hybrid\textsubscript{CPL}. Figs. \ref{fig:increments}A and \ref{fig:increments}B show the mean March--July SIC increments from the 36-year ice-ocean DA simulation and the fully-coupled DA simulation, respectively. Here we can see that the increments show overall very similar magnitudes and spatial patterns, highlighting that the ice-ocean and coupled model have similar sea ice biases in the Arctic. When we then look at the March-initialized reforecasts, we can see the ML increments from Hybrid\textsubscript{IO} are generally negative within the Arctic basin and have larger magnitudes than the ice-ocean DA experiment (Fig. \ref{fig:increments}C vs \ref{fig:increments}A). Meanwhile, the increments from Hybrid\textsubscript{CPL} are in good agreement with the nudged DA experiment (Fig. \ref{fig:increments}D vs \ref{fig:increments}B). Diagnosing each of the inputs to the ML models reveals that sea-surface salinity (SSS) may be causing an out-of-sample issue for Hybrid\textsubscript{IO} (Figs. \ref{fig:increments}E and \ref{fig:increments}F). This is because the ice-ocean DA experiment also includes a restoring of SSS to a monthly climatology, whereas the SSS is allowed to evolve freely in the coupled DA experiment. Therefore, normalizing SSS during Hybrid\textsubscript{IO} reforecasts based on the statistics of the ice-ocean DA experiment produces SSS values $>$4$\sigma$ lower than Hybrid\textsubscript{CPL} in places such as Hudson Bay and the Eurasian coastal seas and $\sim$0.5$\sigma$ lower across the Arctic basin. This highlights that SPEAR generally has a fresher ocean surface than the SSS-restored ice-ocean model.

Conducting the same analysis between June--August for the Antarctic reveals different increment spatial patterns between the two DA experiments (Figs. \ref{fig:increments}G and \ref{fig:increments}H). These differences are most notable in the Weddell Sea, where the ice-ocean DA increments are slightly negative within the interior ice pack, while the coupled DA increments are positive; this emphasizes different sea ice biases between the ice-ocean and coupled model. The Hybrid\textsubscript{IO} increments are then systematically negative in the Weddell Sea and are over 2$\times$ larger in magnitude than those from DA (compare Figs. \ref{fig:increments}I and \ref{fig:increments}G, noting that I has been scaled by 1/2). Meanwhile, the increments from Hybrid\textsubscript{CPL} are more in-line with those from the coupled DA experiment, although show lower magnitude positive increments in the Weddell and Ross seas and larger magnitude negative increments in the marginal ice zone (Fig. \ref{fig:increments}J vs \ref{fig:increments}H). This time, the normalized SSS fields between Hybrid\textsubscript{IO} and Hybrid\textsubscript{CPL} are very similar and generally ``in sample", and the largest differences in ML inputs occur in sea ice thickness (SIT; Fig. \ref{fig:increments}K vs \ref{fig:increments}L). Between June--August, the mean pan-Antarctic sea ice extent in Hybrid\textsubscript{IO} is $\sim$7\% lower than the respective Hybrid\textsubscript{CPL} mean extent. However, the mean pan-Antarctic SIT in Hybrid\textsubscript{IO} is over 25\% lower than Hybrid\textsubscript{CPL}. 

\begin{figure}[t!] 
	\centering
	\includegraphics[width=0.38\textwidth]{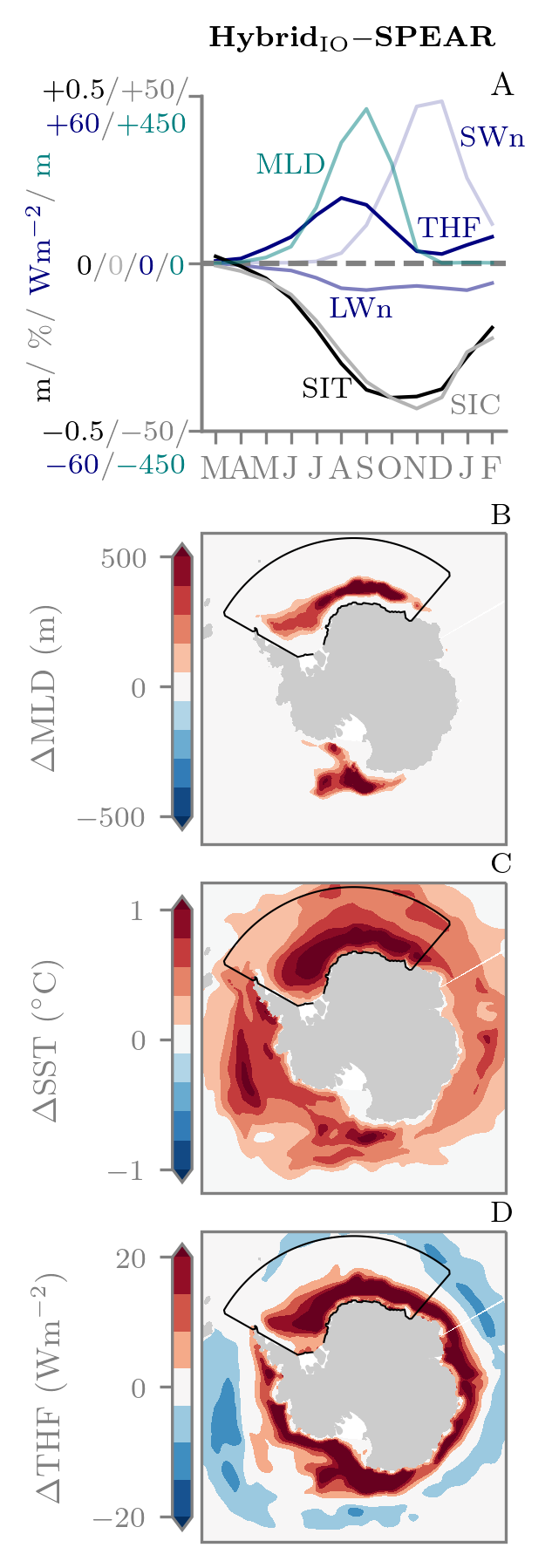}
	\caption{\textbf{March-initialized Hybrid\textsubscript{IO}--SPEAR anomalies, 2018--2024.}
		(\textbf{A}) Mean Weddell Sea anomalies in sea ice concentration (SIC), sea ice thickness (SIT), net shortwave radiation (SWn), net longwave radiation (LWn), turbulent heat flux (THF), and mixed-layer depth (MLD). THF sign convention is positive up, while LW and SW are positive down. (\textbf{B}--\textbf{D}). Average Hybrid\textsubscript{IO} anomalies in MLD, SST, and THF across the 1-year reforecasts. Contour shows region of anomalies in (\textbf{A}).}
	\label{fig:MarchProc} 
\end{figure}

At this point it is worth noting that the ML models in this study do not predict SIT increments, but rather make changes to the model's SIT by adjusting the concentration of ice within each of the model's ice thickness categories (see Materials and Methods). A question therefore remains as to whether Hybrid\textsubscript{IO}'s thinner and less extensive ice between June--August is coming directly from the ML model's SIC updates, or whether the ML model is also triggering feedbacks which inhibit winter ice growth rates and ultimately lead to near ice-free conditions by the end of summer. To investigate this further we take a process-oriented approach by looking at anomalies in coupled ice-atmosphere-ocean diagnostics relative to SPEAR. Fig. \ref{fig:MarchProc}A shows mean Weddell Sea (48.5$^\circ$W--39.5$^\circ$E, 56.61$^\circ$S--90$^\circ$S) anomalies in SIC, SIT, mixed-layer depth (MLD), and surface energy balance terms, for each month of the March-initialized reforecasts. Note that the surface energy balance corresponds to the sum of net shortwave (SWn), net longwave (LWn), and turbulent heat fluxes (THF), where THF are the sum of sensible and latent heat exchanges. THF are also defined as positive up, while LWn and SWn are positive down. For this region of the Weddell Sea, the negative SIC and SIT anomalies indicate an overall negative sea ice volume anomaly relative to SPEAR between March--August. This sea ice volume anomaly is accompanied by a deepening of the ocean mixed layer ($\sim$300 m), as well as an increase in both THF ($\sim$20 Wm$^{-2}$) and upward longwave ($\sim$10 Wm$^{-2}$). This can be explained by the volume anomaly creating areas of open water and thinning the sea ice, both of which make the ocean more susceptible to surface forcing from the atmosphere. This cold wintertime forcing then drives surface cooling and ocean convection (Fig. \ref{fig:MarchProc}B), which brings relatively warm and saline waters to the surface (Fig. \ref{fig:MarchProc}C) and increases THF (Fig. \ref{fig:MarchProc}D)---all of which would inhibit winter ice growth rates. Between March--August solar insolation is also at its lowest, resulting in little to no response from SWn. However, by the time shortwave ``turns on'' in September, the volume anomaly has already had a dramatic impact on the surface albedo. A positive SWn anomaly then grows between September--November and coincides with higher rates of ice loss in Hybrid\textsubscript{IO} (compare Hybrid\textsubscript{IO} and SPEAR sea ice extent curves in Fig. \ref{fig:MarchBias}E). This indicates that the wintertime ocean preconditioning of the sea ice is also potentially triggering summertime ice-albedo feedbacks, further enhancing the sea ice anomaly. The reason that the SIC and SIT anomalies in Fig. \ref{fig:MarchProc}A then start to recover between December--February is because Hybrid\textsubscript{IO} has almost completely lost its ice cover.

\begin{figure}[t!] 
	\centering
	\includegraphics[width=0.5\textwidth]{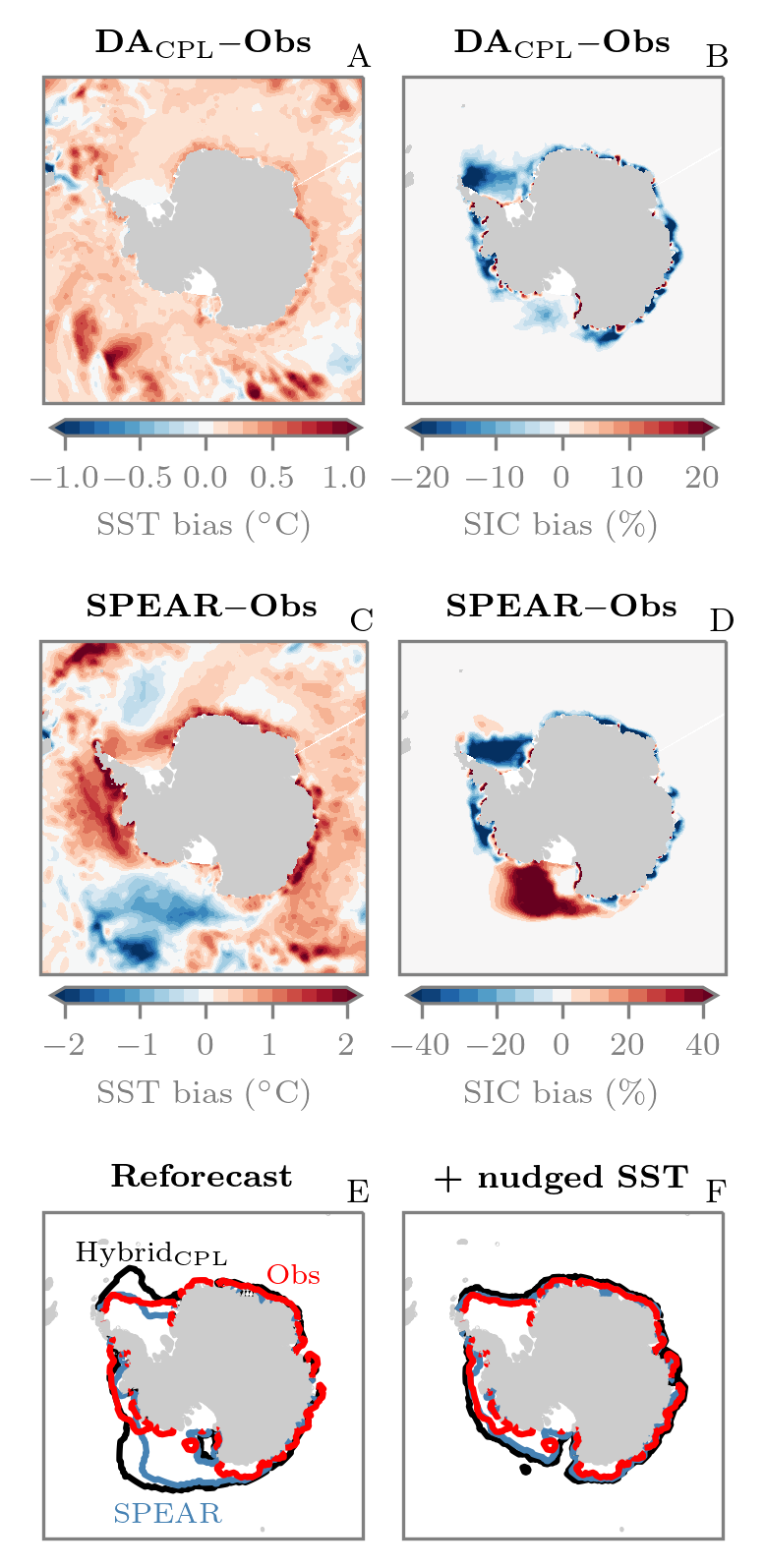}
	\caption{\textbf{February-mean Antarctic sea ice and ocean biases.}
		(\textbf{A},\textbf{B}) 36-year (1982--2017) sea-surface temperature (SST) and sea ice concentration (SIC) biases from the coupled DA experiment, respectively. SST bias relative to Optimum Interpolation SST data \cite{Banzon2016} and SIC bias relative to NSIDC NASA Team observations \cite{DiGirolamo2022}. (\textbf{C},\textbf{D}) Same as (\textbf{A},\textbf{B}) but for November-initialized SPEAR reforecasts of February, 2018--2024. (\textbf{E},\textbf{F}) February sea ice edge locations from November initialized reforecasts without and with SST nudging, respectively.}
	\label{fig:RossBias} 
\end{figure}

Antarctic reforecasts with Hybrid\textsubscript{IO} appear to be an example of how interactions between ML models and climate physics can cause out-of-sample behavior and potential runaway feedbacks. In this case, this occurs by the ML model pre-conditioning the winter sea ice and ocean state to facilitate near ice-free conditions by end of summer. Evaluating the same Fig. \ref{fig:MarchProc} diagnostics for Hybrid\textsubscript{CPL} reveals a stable simulation with no sizable anomalies relative to SPEAR (not shown), highlighting that ice-atmosphere-ocean feedbacks within ML training data are essential for online generalization in coupled models. Finally, evaluating Hybrid\textsubscript{IO} in the Arctic also does not show the same pathological behavior as the Antarctic, with mean anomalies in surface energy balance terms on the order of 1 Wm$^{-2}$ and MLD anomalies of less than 5 m across the Arctic basin (see Fig. \ref{fig:MarchProcArctic}). This may indicate that an ML model trained in an ice-ocean configuration could generalize to the fully-coupled SPEAR model in the Arctic after careful treatment of nudging routines, such as SSS.

\subsection*{Impact of coupled model biases on ML generalization}

At this point we have established that Hybrid\textsubscript{CPL} is the desirable hybrid model for global sea ice bias correction. However, in Fig. \ref{fig:forecastskill}F we saw that Hybrid\textsubscript{CPL} also shows degradations in forecast skill relative to SPEAR for target months in Antarctic summer (January--April). We hypothesize in this section that these degradations originate from an out-of-sample problem related to coupled model biases. Recall that the Hybrid\textsubscript{CPL} ML model was trained on model state variables which were generated from a simulation which performs SIC DA as well as SST and atmospheric nudging. We have then implemented this ML model into reforecast experiments with a free-running atmosphere and ocean. Learning DA increments in this nudged configuration was intended to create an environment in which the ML model learns intrinsic sea ice model physics errors, as opposed to coupled model biases which imprint on the sea ice. However, if the ML model has not been exposed to these biases, then it could make erroneous online predictions. In Fig. \ref{fig:RossBias}A we can see that the 36-year coupled DA simulation (which performs SST nudging) contains a slight positive summertime (February) SST bias. The resultant February SIC from this simulation (Fig. \ref{fig:RossBias}B) also has an Antarctic-wide low bias---the February SIC DA increments will therefore be positive to counteract this bias. In the free-running coupled reforecasts, SPEAR exhibits larger and more heterogeneous February SST biases (Fig. \ref{fig:RossBias}C; we use November-initialized forecasts of February as an example here, but the same relationship holds for other initialization dates and summer target months). One noteworthy region is the Ross Sea, which contains a large area of negative SST bias. In Fig. \ref{fig:RossBias}D we then see this SST bias imprinted onto the sea ice as a positive SIC bias. Based on the DA simulation, the ML model has learned to add sea ice in Antarctic summer. However, in the online reforecasts it is now adding sea ice onto a pre-existing positive Ross Sea bias. We can see this in Fig. \ref{fig:RossBias}E, which shows the February sea ice edge contour for each reforecast experiment, highlighting the fact that Hybrid\textsubscript{CPL} has exacerbated the sea ice bias in the Ross Sea. Furthermore, Hybrid\textsubscript{CPL} has also exacerbated a slight positive bias in the Weddell Sea, which may also be related to the negative SST bias in this location (see Fig. \ref{fig:RossBias}C). We further test our hypothesis by repeating the November-initialized reforecasts, but this time with SST nudging turned on. In Fig. \ref{fig:RossBias}F we can see that Hybrid\textsubscript{CPL} performs better in this scenario, with a sea ice edge that is in closer agreement with observations and SPEAR in the Ross and Weddell seas.

\begin{figure}[t!] 
	\centering
	\includegraphics[width=\textwidth]{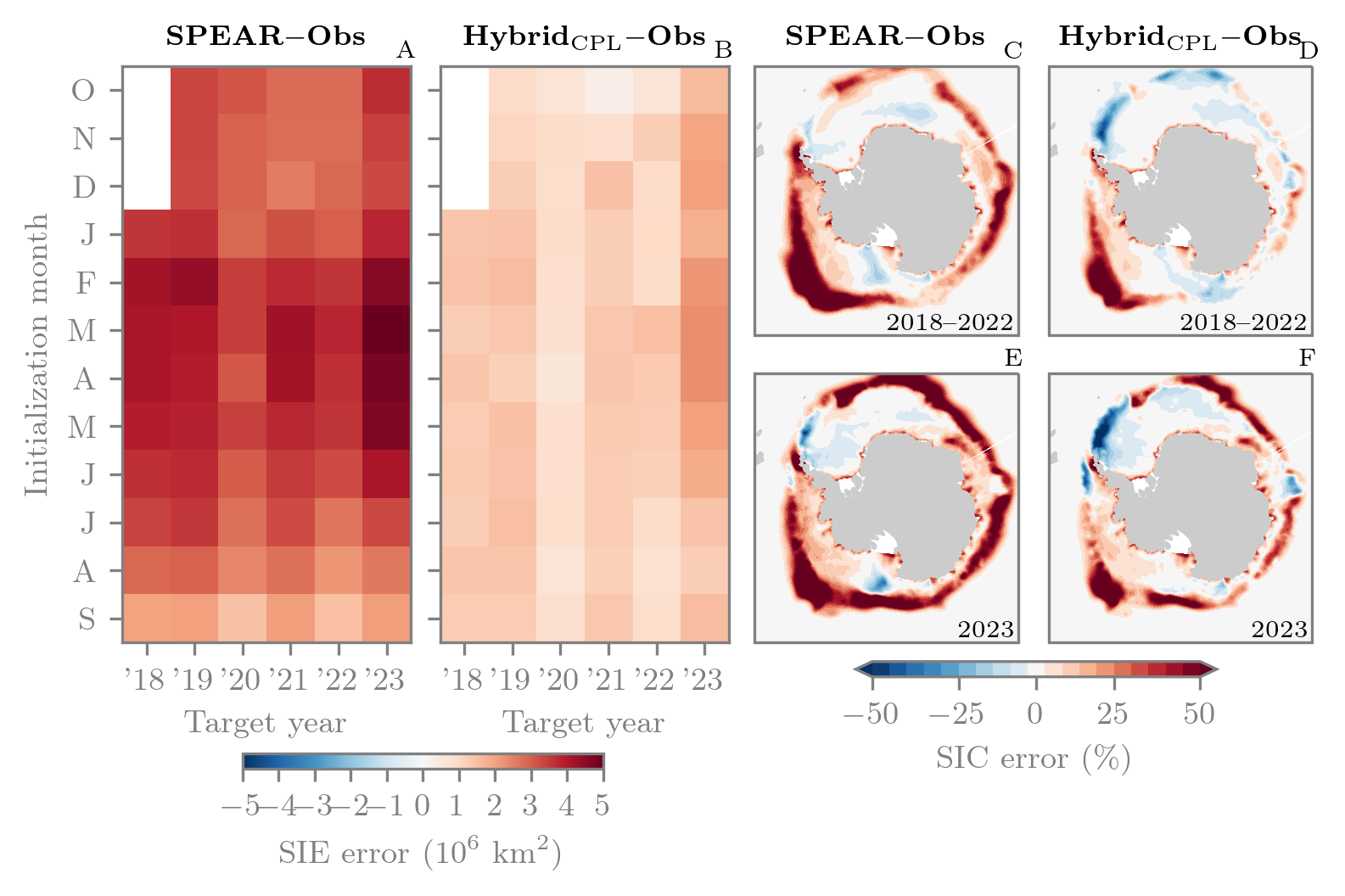}
	\caption{\textbf{September Antarctic sea ice prediction error.}
		(\textbf{A},\textbf{B}) Pan-Antarctic September sea ice extent (SIE) error for each year between 2018--2023, for SPEAR and Hybrid\textsubscript{CPL}, respectively. (\textbf{C},\textbf{D}) Mean September sea ice concentration (SIC) error for March--May initialized reforecasts, for SPEAR and Hybrid\textsubscript{CPL}, respectively. (\textbf{E},\textbf{F}) Same as (\textbf{C},\textbf{D}) but for 2023. All errors are relative to NSIDC NASA Team observations \cite{DiGirolamo2022}.}
	\label{fig:SepErr} 
\end{figure}

\subsection*{Extreme events: September 2023 Antarctic case study}
With just 7 years of validation data, we cannot robustly assess the ability of Hybrid\textsubscript{CPL} to predict sea ice anomalies through typical metrics such as de-trended anomaly correlation coefficient. Instead, we briefly look at the September 2023 record low sea ice maximum in the Antarctic to determine the \emph{potential} for hybrid models to yield improved forecasts in extreme years. The 2023 September Antarctic sea ice extent gained considerable attention for being a ``once in a multi-million-year event'' with an extent anomaly $>$5$\sigma$ below the 1980--2010 mean \cite{Gilbert2024}. This anomaly was caused in large part due to anomalously warm upper ocean temperatures and strong northerly winds, both of which significantly inhibited winter ice growth rates in the Ross and Weddell seas \cite{Jena2024}.

In Fig. \ref{fig:SepErr} we show the September Antarctic sea ice prediction skill for both SPEAR and Hybrid\textsubscript{CPL}. In terms of pan-Antarctic sea ice extent, SPEAR has a positive September extent bias for all initialization months and years (Fig. \ref{fig:SepErr}A), where the bias grows steadily with increasing lead time, up to approximately February. The largest September 2023 forecast errors occur for initialization dates March--May, with an average extent bias of 4.77 million km$^2$. Meanwhile, the average 2018--2022 extent bias for March--May forecasts is 3.80 million km$^2$---a 0.97 million km$^2$ error increase from 2018--2022 to 2023. The September sea ice extent bias for Hybrid\textsubscript{CPL} is then systematically lower than SPEAR, for all initialization months and years (Fig. \ref{fig:SepErr}B). Interestingly, for years 2018--2022, the forecast error does not grow with lead time at the same rate as SPEAR. For example, the difference in sea ice extent bias for February-initialized forecasts vs September-initialized is just 0.08 million km$^2$, while for SPEAR it is 2.05 million km$^2$. For March--May initialized forecasts in 2023, Hybrid\textsubscript{CPL} shows a 1.12 million km$^2$ increase in forecast error compared to 2018--2022, from 1.21 to 2.33 million km$^2$. While the magnitude of this error increase is relatively similar for Hybrid\textsubscript{CPL} and SPEAR (1.12 vs 0.97, respectively), the absolute error for March--May forecasts with Hybrid\textsubscript{CPL} is still $>$2$\times$ lower than SPEAR. 

Figs. \ref{fig:SepErr}C and \ref{fig:SepErr}D show the average September SIC error for March--May initialized forecasts between 2018--2022, for SPEAR and Hybrid\textsubscript{CPL}, respectively. This shows that the hybrid model is removing a significant amount of error along the ice edge and is removing some of the large SIC bias in the Ross and Amundsen seas. Comparing these figures to September 2023 (Figs. \ref{fig:SepErr}E and \ref{fig:SepErr}F), we see that SPEAR has larger ice edge errors than compared to 2018--2022, particularly in the Weddell and Ross seas, as well as the Pacific sector. SPEAR also has a localized negative SIC error in the Ross Sea ice pack. Hybrid\textsubscript{CPL} shows increased ice edge errors relative to its 2018--2022 counterpart. The Hybrid\textsubscript{CPL} error pattern generally resembles a muted version of the SPEAR error pattern, except for the Weddell Sea, where the errors are exacerbated, and the Ross Sea, where the negative error in SPEAR is no longer present.

This September Antarctic case study demonstrates a successful example of how hybrid models can systematically improve seasonal sea ice forecasts. However, due to the fact that the error increase between 2018--2022 and 2023 is roughly consistent for SPEAR and Hybrid\textsubscript{CPL}, we cannot confidently say here that Hybrid\textsubscript{CPL} is better equipped to capture extreme events. Despite this, the systematic bias improvements from Hybrid\textsubscript{CPL} does suggest an improvement in the ``quality'' of our ensemble forecast system, where quality can be quantified in terms of the ratio of the ensemble mean forecast error (RMSE) to the spread (1$\sigma$) in the ensemble---the so-called spread-skill metric. This ratio should be approximately equal to 1 for a well-behaved model, while for SPEAR it is on average equal to 10 for September Antarctic sea ice forecasts (see Fig. \ref{fig:SpreadSkill}). While the average ratio is significantly improved for Hybrid\textsubscript{CPL} at 3.4, the model is still considerably under-dispersive, meaning that it may still struggle to capture extreme events within its forecast ensemble. Nevertheless, the improvements in sea ice mean state suggest that Hybrid\textsubscript{CPL} can potentially improve the representation of coupled sea ice processes in seasonal forecasts, such as Southern Ocean net primary productivity \cite{Holland2025} or surface air temperature \cite{Guemas2016b}. This goes beyond the scope of our present study, although will be investigated in future work.

\section*{Discussion}

This study introduced a hybrid modeling framework which uses ML to bias correct global sea ice conditions during a set of 1-year fully-coupled forecast experiments with the GFDL SPEAR climate model. The ML models in this study were trained to predict SIC DA increments using only information from local model state variables, yielding a state-dependent representation of the sea ice model errors. We have paid particular attention to how training ML models on DA increments generated from reanalysis-forced vs nudged configurations of SPEAR are able to generalize to the fully-coupled free-running SPEAR model. We referred to the two resultant hybrid models from these training configurations as Hybrid\textsubscript{IO} and Hybrid\textsubscript{CPL}, respectively.

Reforecast experiments initialized between 2018--2023 show that Hybrid\textsubscript{CPL} outperforms SPEAR in the Arctic for all target months other than October and November, for which there are only marginal degradations in pan-Arctic RMSE of SIC ($<$1\%). Meanwhile,  Hybrid\textsubscript{IO} shows systematic degradations relative to SPEAR for target months May--November ($\sim$4.5\% increase in SIC RMSE), which is due to out-of-sample behavior originating from ML input variables, particularly SSS. In the Antarctic, Hybrid\textsubscript{IO} also systematically degrades SPEAR forecasts between July--January ($\sim$10\% increase in RMSE). This is due to a combination of out-of-sample behavior and coupled feedbacks between the ML model and physical processes within SPEAR. For one, the mean Antarctic DA increments show different spatial patterns between the reanalysis-forced and nudged fully-coupled model, highlighting that these two model configurations have different sea ice biases. Therefore, learning increments in the ice-ocean model does not generalize to the fully-coupled model. Furthermore, the Antarctic reforecasts with Hybrid\textsubscript{IO} trigger a sequence of coupled feedbacks whereby the ML model first creates negative sea ice concentration and thickness anomalies relative to SPEAR. This then increases ocean vertical mixing, which brings more heat to the surface and further exacerbates the negative volume anomaly. These processes significantly impact the sea ice mean state in Hybrid\textsubscript{IO}, with a sea ice wintertime maximum occurring one month earlier than in SPEAR and summertime conditions which are practically ice free. Conversely, the Hybrid\textsubscript{CPL} configuration systematically outperforms SPEAR between May--December, reducing the September Antarctic forecast bias by more than a factor of 2. While our relatively short validation period has not allowed us to confidently assess whether Hybrid\textsubscript{CPL} is more skillful at predicting sea ice anomalies, we hypothesize that a significantly improved mean state will inherently allow the forecast ensemble to capture a more realistic range of events, as seen in other bias correction studies \cite{Lu2020}. Nevertheless, an extension of the present methodology could be to train on anomaly increments, which has shown success at improving the representation of large-scale atmospheric modes of variability \cite{Chapman2025}.

The improved online generalization with Hybrid\textsubscript{CPL} underscores a central takeaway from our study, that exposing ML models to coupled ice-atmosphere-ocean processes is essential for robust online performance in free-running coupled model simulations. Our framework therefore provides a promising step towards improving operational numerical seasonal predictions with ML. However, achieving this goal first requires attention of Hybrid\textsubscript{CPL}'s forecast degradations in Antarctic summer. We showed that these degradations are likely originating from coupled ocean model biases, which were not present in the training data due to the DA simulation containing SST nudging. While online generalization could potentially be improved by generating the sea ice DA increments in a free-running configuration of SPEAR, we have endeavored to remain consistent with past studies \cite{Gregory2023,Gregory2024a}, which constrained the sea ice in this way to target intrinsic sea ice model physics errors. Directly targeting sea ice model physics errors allows for flexibility in terms of future model development, whereby the learned errors can be potentially attributed to specific deficiencies within pre-existing parameterization schemes \cite{Rodwell2007}. In any case, we have shown that applying SST nudging on top of our ML-based bias correction in these 1-year reforecast experiments significantly improves online generalization in Antarctic summer. Therefore, future work will involve running weakly-coupled DA experiments where assimilation is performed in both the sea ice and ocean components. This will provide a consistent set of ocean and sea ice increments with which to train ML models and apply together during subsequent reforecast experiments. 

Finally, while we have showcased our ML-based bias correction framework in 1-year reforecasts here, the methodology also has potential for climate-timescale integrations. Achieving this requires further development of the methodology towards a conservative implementation of the corrections. At present, the sea ice increments are applied to the SIC state at every thermodynamic timestep by simply adding or removing sea ice within a given grid cell. In reality, these updates should also make changes to the heat, water mass, and salt content of the ocean mixed layer. Conserving heat poses significant challenges and warrants investigation. One past study showed that a conservative ocean temperature tendency adjustment approach can be achieved by ensuring that the global integral of the temperature corrections equals zero \cite{Lu2020}. This is likely insufficient for our sea ice case, where we often need to make a net change to the sea ice state. However, our ML framework could be updated to conserve water mass and salt by computing an appropriate surface heat flux (q-flux) which would create the necessary SIC change predicted by the ML model---an approach which has been proposed for sea ice nudging during polar amplification model intercomparison project simulations \cite{Sun2020}. Conserving the water mass budget would be crucial for understanding how such an ML scheme impacts large-scale overturning circulation patterns in the ocean, for example.

\section*{Materials and Methods}
\subsection*{The GFDL SPEAR model}
\noindent The Seamless system for Prediction and EArth system Research (SPEAR) is a fully-coupled ice-atmosphere-ocean-land model \cite{Delworth2020}. There are two configurations of SPEAR that are routinely run at GFDL for climate simulations and seasonal predictions: SPEAR\textsubscript{LO} and SPEAR\textsubscript{MED}. These two configurations differ only in the horizontal resolution of their atmospheric and land components, at 1$^\circ$ and 0.5$^\circ$, respectively. Otherwise, both configurations contain 33 vertical levels in the atmosphere, 75 vertical levels in the ocean, with the atmosphere, land, ocean and sea ice based on AM4.0, LM4.0, MOM6 and SIS2, respectively \cite{Zhao2018a,Zhao2018b,Adcroft2019}. The ocean and sea ice components are configured to a nominal 1$^\circ$ horizontal resolution in both SPEAR\textsubscript{LO} and SPEAR\textsubscript{MED}. Although SPEAR\textsubscript{MED} generally outperforms SPEAR\textsubscript{LO} in terms of seasonal Antarctic sea ice forecasts \cite{Bushuk2021}, our study focuses on the relative improvements of a given climate model's sea ice forecasts through our hybrid ML scheme. We therefore opt for SPEAR\textsubscript{LO} (hereafter SPEAR) given its computational advantage.

\subsection*{Generating the training data}
For details of the reanalysis-forced ice-ocean simulation used to train the ML model for Hybrid\textsubscript{IO}, we refer the reader to studies \cite{Gregory2023,Gregory2024a}. The model state variables and DA increments that are used to train the ML model for Hybrid\textsubscript{CPL} are generated from a SPEAR 30-member large ensemble simulation spanning 1982--2017. The initial conditions for this simulation are from a perturbed physics spin-up run off of a SPEAR large ensemble historical simulation spanning 1851--2010. Specifically, we re-run the historical large ensemble simulation between January 1 1968 and January 1 1979, but with perturbed sea ice physics parameters for each ensemble member (see \cite{Gregory2023} for details of these perturbations). Then from January 1 1979 to January 1 1982 the 3D atmospheric temperature, winds and humidity fields are nudged to the NOAA climate forecast system reanalysis (CFSR; \cite{Saha2010}) at a 6-hourly e-folding timescale for temperature and winds, and 24-hourly for humidity. From January 1 1982 to January 1 2018 we then nudge SSTs toward version 2.0 of the NOAA optimum interpolation SST (OISST) product \cite{Banzon2016} at a piston velocity of 4 meters per day, which corresponds to a timescale of 12.5 days for a 50-meter mixed layer. We also nudge the atmosphere to CFSR as before and assimilate NASA Team passive microwave SIC observations from the National Snow and Ice Data Center \cite{DiGirolamo2022} into SIS2 using the ensemble adjustment Kalman filter (EAKF; \cite{Anderson2001}). It should be noted that sea ice covered grid points within the raw OISST data are assigned a fixed value of $-1.8^\circ$C. During nudging, we then replace OISST values of $-1.8^\circ$C with the salinity-dependent freezing point of sea water ($\mathrm{T}_f$) at each timestep, based on the model's SSS and the model's empirical freezing-point equation $\mathrm{T}_f = -0.054\mathrm{SSS}$. Without this change the SST nudging can trigger spurious ice-growth feedbacks in regions of fresh water such as the East Siberian and Laptev seas, which have freezing points that are warmer than $-1.8^\circ$C. We also note here that SIS2 has a 5-category sub-grid sea ice thickness distribution, where the aggregate, or observable, SIC is given as the sum of the concentration in each category (see \cite{Gregory2023} for further details). Providing observations are available, sea ice DA is performed every 5 days over the course of the 36-year simulation. From this simulation, we then compute the 5-day mean of all model state variables, providing 2619 pairs of model state variables (inputs) and DA increments (outputs) to train the ML model.

\subsection*{Machine learning model and implementation}
The ML framework proposed by \cite{Gregory2023} uses a CNN to map model state variables and their tendencies to the aggregate SIC increment from DA (the sum of the increments in each sub-grid thickness category). The input variables for this CNN are SIC, SST, zonal and meridional components of ice velocities, SIT, SWn, ice-surface skin temperature, SSS, and finally a land-sea mask (17 inputs in total). The predicted increment from this CNN is then passed to an artificial neural network (ANN), along with state variables and tendencies corresponding to the sub-grid category SIC fields and a land-sea mask, to predict the SIC DA increments of each category. In \cite{Gregory2024a}, this ML architecture was then used to bias correct ice-ocean simulations every 5 days across a 5-year simulation. While this approach systematically reduced global sea ice biases, it left egregious sawtooth-type imprints of the 5-day corrections in the resultant simulation. In Fig. \ref{fig:IORestarts} we show that increasing the frequency of the ML corrections to 2 days in this same ice-ocean configuration (and linearly scaling the predicted increments by 2/5), leads to poor performance. This is due to out-of-sample issues related to the model state tendencies (not shown). However, by removing the tendencies from the list of inputs and re-training the networks we achieve stable online performance at 1-day implementation frequency (Fig. \ref{fig:IORestarts}C). This 1-day implementation subsequently removes all correction imprints. We therefore use these same subsets of inputs for both Hybrid\textsubscript{IO} and Hybrid\textsubscript{CPL} in our present study. Specifically, the CNN uses 9 inputs and the ANN uses 7 inputs (see above). Note that both ML models for Hybrid\textsubscript{IO} and Hybrid\textsubscript{CPL} have also gone through a refinement procedure in which the training data are iteratively augmented by running new simulations that apply ML and then DA corrections every 5 days. This has been shown to dramatically improve online performance of ML-based bias correction methods \cite{Gregory2024a}.

The 1-day implementation in  Fig. \ref{fig:IORestarts}C is achieved by performing offline updates to the model restart files in Python, which has a $\sim$440\% slowdown cost associated with pausing and restarting the model every day. To address this issue, we implement the ML models directly into the SIS2 source code and apply the corrections to SPEAR reforecasts at the sea ice thermodynamic timestep (30 minutes). The CNN and ANN architectures are relatively simple, consisting only of 2D convolution operations, local weighted sums, and ReLU functions. We therefore also code these directly into Fortran, rather than relying on a Fortran-Python wrapper such as~\textsc{FTorch} to do the inference \cite{Atkinson2025}. 

While developing this approach, we initially encountered generalization issues related to the fact that the CNN has been trained on 5-day-mean input fields, which smooths out features including the diurnal cycle and sharp gradients associated with sub-5-daily variability---features which are prevalent in SWn, ice velocities, and surface skin temperature instantaneous fields (see Fig. \ref{fig:meanInputs}). We address this issue through a pragmatic solution of gathering the network inputs over the first day of the simulation to compute a daily mean. With these daily-mean fields, we then do inference with the ML model at 00:00 UTC and apply this predicted correction to the category SIC states at every timestep over the course of the proceeding day (while also accumulating the network inputs again for the next daily-mean computation). Note that we also scale the predicted increment by 1/240 to account for a 30-minute thermodynamic timestep. This procedure then continues for the length of the simulation. Through this configuration, the network receives the same input fields as the offline restart approach, although now spreads the corrections across each timestep. This way, the hybrid approach maintains roughly equivalent throughput as the free-running SPEAR model (0.3\% slowdown), even when doing ML inference on CPU. Finally, in the case where sea ice is added to a grid cell which was previously ice-free, we assign this new ice a thickness of 0.05, 0.2, 0.5, 0.9, 2.0 meters for sub-grid categories 1 through 5, respectively. We also assign the new ice a salinity of 5 psu and a temperature of $-2^\circ$C.

\subsection*{Reforecast initialization procedure}
The initial conditions are identical for the SPEAR and Hybrid reforecasts and are based on a series of ocean and sea ice DA experiments. For the ocean, initial conditions come from a 30-member SPEAR ocean DA simulation spanning 1990--2023, within which NOAA OISST data, Argo temperature and salinity floats, expendable bathythermograph data and tropical moorings are assimilated daily using the EAKF \cite{Lu2020}. The sea ice, atmosphere and land initial conditions for both SPEAR and Hybrid reforecasts correspond to simply extending the 1982--2017 sea ice DA simulation that was used to generate the ML training data, from January 1 2018 to July 1 2023.

The SPEAR and Hybrid reforecasts in this study are configured as 15-member ensemble experiments which run for 1 year. This corresponds to combining the first 15 members of the atmosphere, land, and sea ice initial conditions from the sea ice DA experiment with the first 15 members of the ocean DA experiment. The reforecasts also include an ``ocean tendency adjustment'' approach, which applies a climatology correction to the 3D ocean temperature and salinity fields on month of the year \cite{Lu2020}. This approach has been shown to reduce ocean model bias in climate simulations with SPEAR and also improve the seasonal prediction skill of El Ni\~no Southern Oscillation.








\clearpage 

%
\bibliography{science_template} 
\bibliographystyle{sciencemag}

%
%
%
%
%
%


\section*{Acknowledgments}
This work was intellectually supported by various other members of the M$^2$LInES project, as well as being supported through the provisions of computational resources from the National Oceanic and Atmospheric Administration (NOAA) Geophysical Fluid Dynamics Laboratory (GFDL). The authors also thank William Cooke and Xiaosong Yang for their technical support, as well as Matthew Harrison and Danni Du for their invaluable feedback on this work.
\paragraph*{Funding:}
This work received support through Schmidt Sciences, LLC.
\paragraph*{Author contributions:}
WG was responsible for development and implementation of the machine learning algorithm in the SPEAR climate model, running the analysis and writing the manuscript. MB provided intellectual support on development and implementation. YZ created the infrastructure for performing sea ice data assimilation in the GFDL sea ice model. AA provided intellectual support and was responsible for conceptualization. LZ provided intellectual support and was responsible for conceptualization. CM and LJ provided technical support and code for running SPEAR simulations on the GFDL high-performance computer. All authors contributed to the development of this manuscript.
\paragraph*{Competing interests:}
There are no competing interests to declare.
\paragraph*{Data and materials availability:}
The SPEAR and Hybrid reforecast data are available on Zenodo https://doi.org/10.5281/zenodo.15343808. This repository also contains the neural network weights, normalization statistics, and training data for Hybrid\textsubscript{CPL}. Furthermore, a GitHub link is provided in this repository which points to the SIS2 code where the ML model is implemented. The training data for Hybrid\textsubscript{IO} can then be found at https://doi.org/10.5281/zenodo.7818178.


\subsection*{Supplementary materials}
Figs. S1 to S5


\newpage


\renewcommand{\thefigure}{S\arabic{figure}}
\renewcommand{\thetable}{S\arabic{table}}
\renewcommand{\theequation}{S\arabic{equation}}
\renewcommand{\thepage}{S\arabic{page}}
\setcounter{figure}{0}
\setcounter{table}{0}
\setcounter{equation}{0}
\setcounter{page}{1} 


\begin{center}
\section*{Supplementary Materials for\\ \scititle}

William~Gregory$^{\ast}$,
Mitchell~Bushuk,
Yong-Fei~Zhang, 
Alistair~Adcroft,
Laure~Zanna,\\ 
Colleen McHugh,
Liwei Jia \\
\small$^\ast$Corresponding author. Email: wg4031@princeton.edu\\
\end{center}

\subsubsection*{This PDF file includes:}
Figs. S1 to S5





\newpage



\begin{figure}[t!] 
	\centering
	\includegraphics[width=\textwidth]{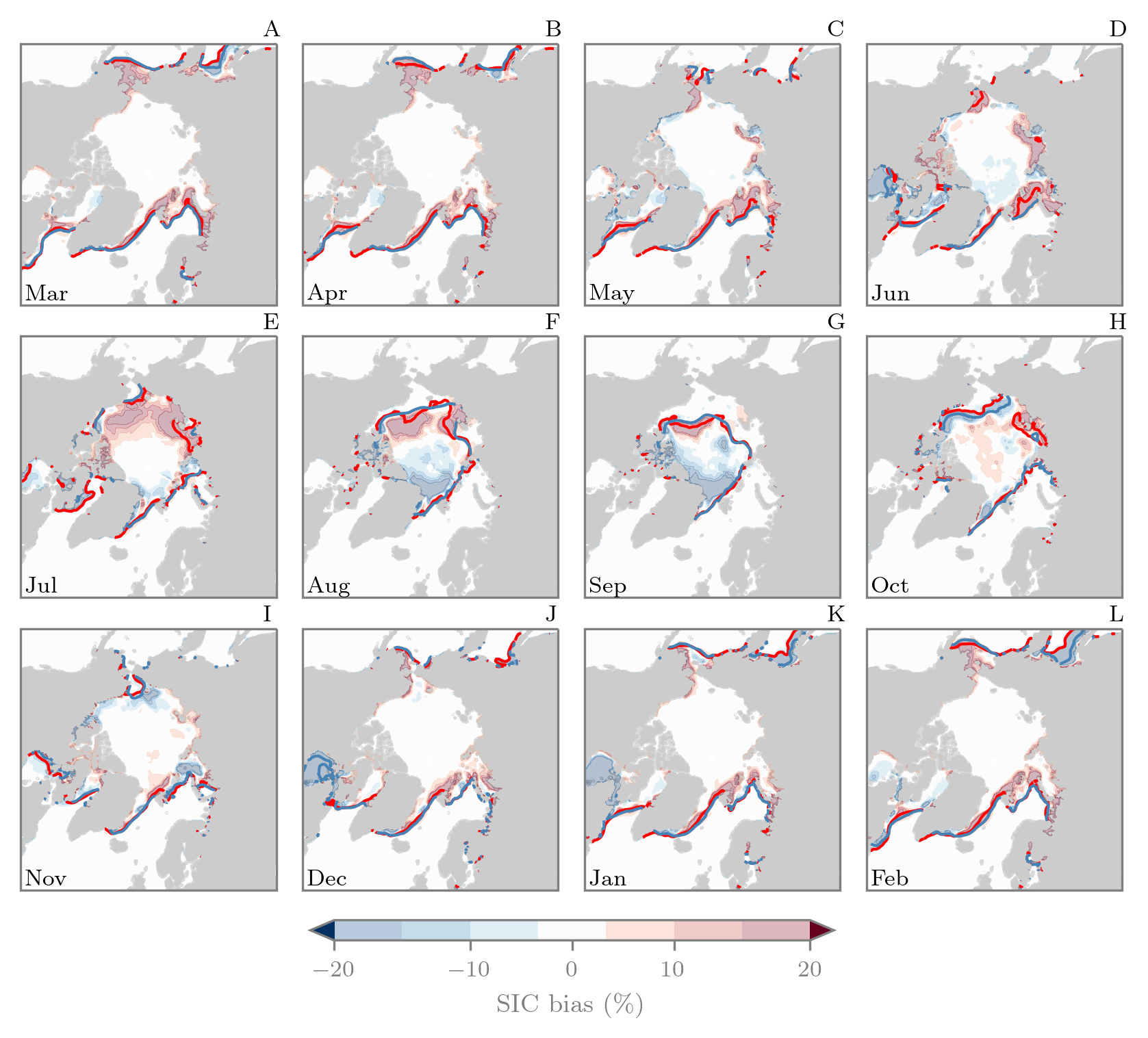}
	\caption{\textbf{March-initialized SPEAR reforecast bias, 2018--2024.}
		(\textbf{A}--\textbf{L}) March--February monthly-mean sea ice concentration (SIC) bias for SPEAR reforecasts, relative to NSIDC NASA Team observations \cite{DiGirolamo2022}, and the associated monthly-mean sea ice edge contour for SPEAR (blue) and observations (red).}
	\label{fig:MarchSPEAR} 
\end{figure}

\begin{figure}[t!] 
	\centering
	\includegraphics[width=0.38\textwidth]{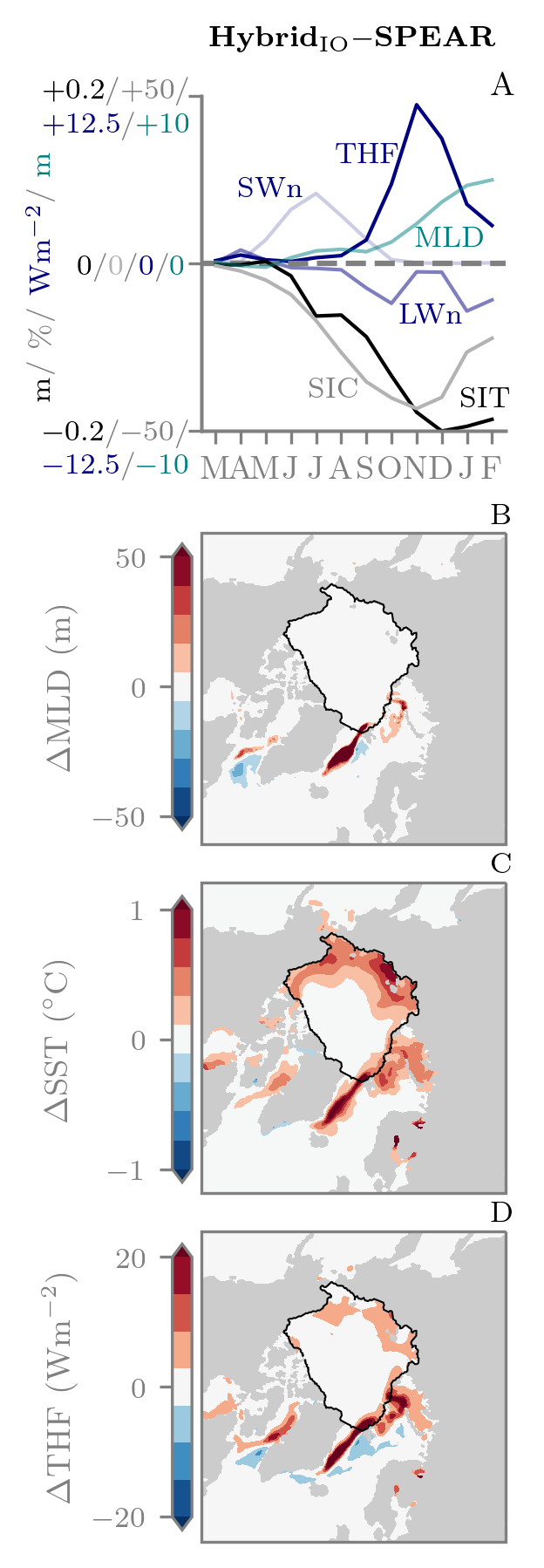}
	\caption{\textbf{March-initialized Hybrid\textsubscript{IO}--SPEAR anomalies, 2018--2024.}
		(\textbf{A}) Mean Arctic basin anomalies in sea ice concentration (SIC), sea ice thickness (SIT), net shortwave radiation (SWn), net longwave radiation (LWn), turbulent heat flux (THF), and mixed-layer depth (MLD), for Hybrid\textsubscript{IO}. THF sign convention is positive up, while LW and SW are positive down. (\textbf{B}--\textbf{D}). Average Hybrid\textsubscript{IO} anomalies in MLD, THF, and LWn across the 1-year reforecasts. Contour shows region of anomalies in (\textbf{A}).}
	\label{fig:MarchProcArctic} 
\end{figure}

\begin{figure}[t!] 
	\centering
	\includegraphics[width=0.65\textwidth]{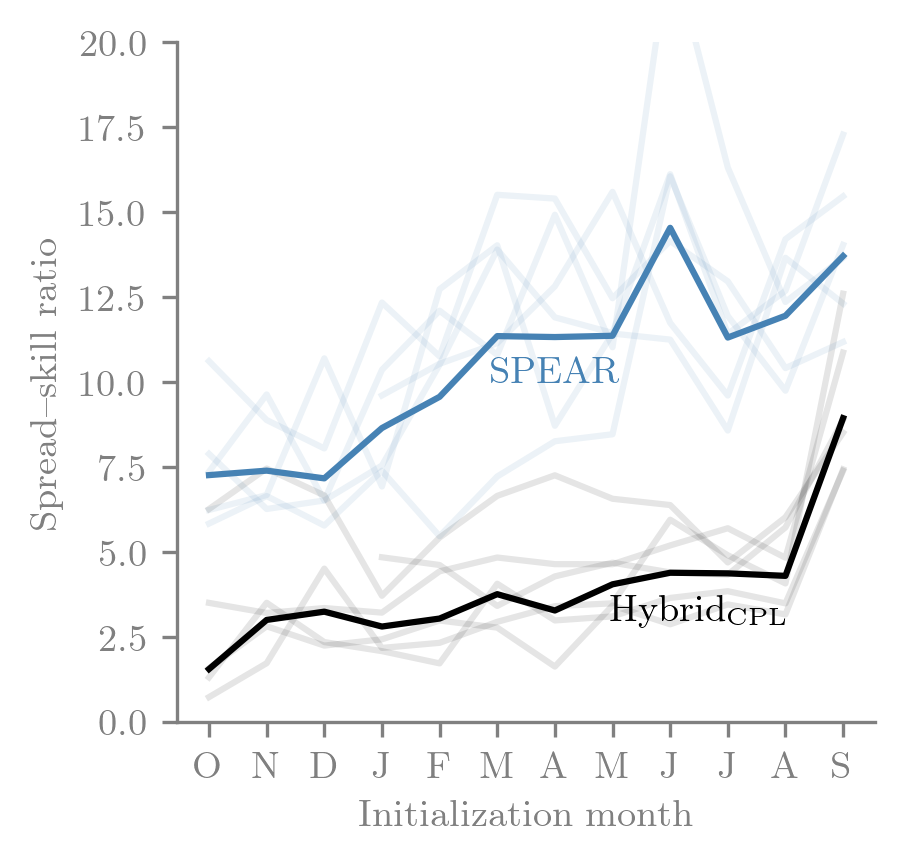}
	\caption{\textbf{Spread--skill ratio for pan-Antarctic September sea ice extent forecasts.}
		Transparent lines are for individual forecast years between 2018--2023, while the opaque lines are the mean across all years. Forecast lead time increases from right to left on x-axis.}
	\label{fig:SpreadSkill} 
\end{figure}

\begin{figure}[t!] 
	\centering
	\includegraphics[width=\textwidth]{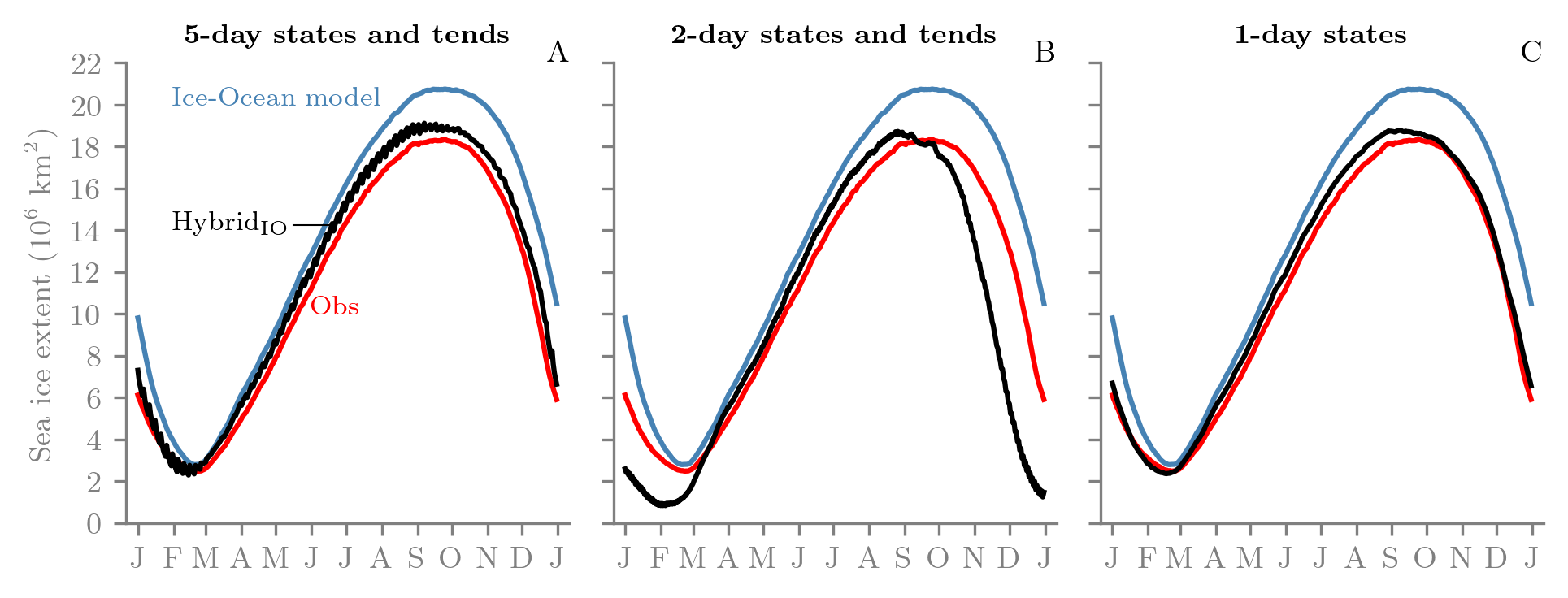}
	\caption{\textbf{Different ML implementation frequencies in ice-ocean simulations, 2018--2022.}
		(\textbf{A}) Mean pan-Antarctic sea ice extent for a reanalysis-forced ice-ocean model (blue), observations (red), and an ice-ocean simulation which performs ML-based bias correctione every 5 days via offline updates to model restarts (black). (\textbf{B}) Using the same ML model as in (\textbf{A}) but doing ML corrections every 2 days. (\textbf{C}) Drop model tendencies from ML inputs and implement every 1 day.}
	\label{fig:IORestarts} 
\end{figure}

\begin{figure}[t!] 
	\centering
	\includegraphics[width=\textwidth]{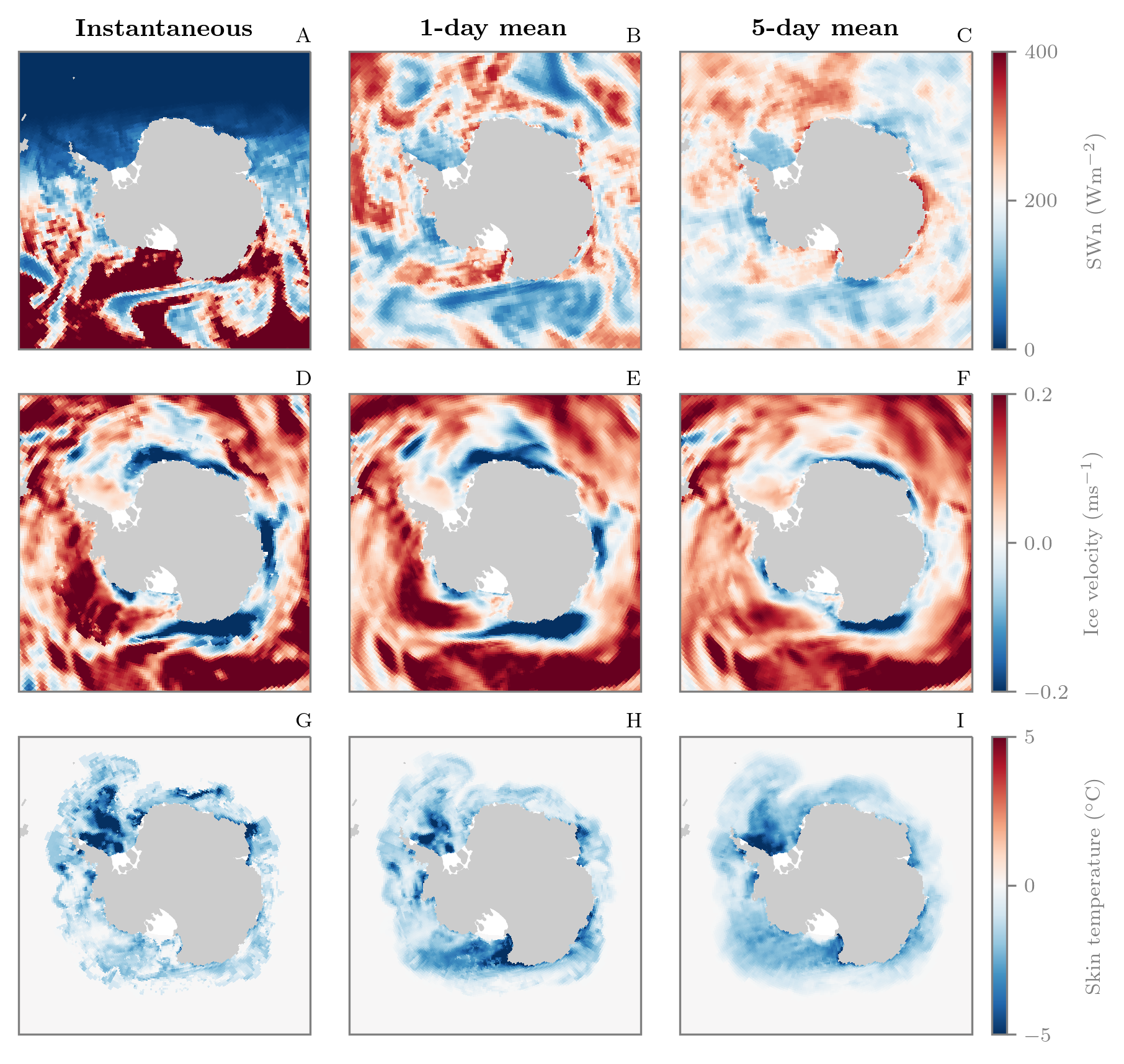}
	\caption{\textbf{Difference in ML inputs across temporal averaging windows.}
		(\textbf{A}) Instantaneous SWn on Januay 1 2018 at 00:30 UTC, (\textbf{B}) the 1-day mean SWn from January 1 2018 00:00 UTC to January 2 2018 00:00 UTC, (\textbf{C}) the 5-day mean SWn from January 1 2018 00:00 UTC to January 6 2018 00:00 UTC. (\textbf{D}--\textbf{F}) Same as (\textbf{A}--\textbf{C}) but for zonal ice velocity. (\textbf{G}--\textbf{I}) Same as (\textbf{A}--\textbf{C}) but for ice-surface skin temperature. Notice features like the diurnal cycle in (\textbf{A}) or grid artifacts in (\textbf{D}) and (\textbf{G}). All ML models in this study are trained on 5-day mean state variables.}
	\label{fig:meanInputs} 
\end{figure}





\end{document}